\documentstyle[preprint,revtex]{aps}
\begin{document}
\draft
\preprint{IMSc 92/09, 25/2/92}
\preprint{hepth@xxx/9203013}
\begin{title}
Is the $O(3)~\sigma$ Model with the Hopf Term\\
Exactly Equivalent to a Higher Spin Theory?
\end{title}
\author
{T.R.Govindarajan and R.Shankar}
\begin{instit}
The Institute of Mathematical Sciences, C.I.T. Campus, Madras-600113, INDIA.
\end{instit}
\author
{N.Shaji}
\begin{instit}
Centre for Development of Imaging Technology, Trivandrum-695027, INDIA.
\end{instit}
\author
{M.Sivakumar}
\begin{instit}
School of Physics, University of Hyderabad, Hyderabad-500134, INDIA.
\end{instit}
\begin{abstract}
We write down a  local $CP_1$ model involving two gauge fields, which is
exactly equivalent to the O(3)  $\sigma$ model with the Hopf term. We
impose the $CP_1$ constraint by using the gaussian representation of the
delta function. For the coefficient of the Hopf term, $\theta = {\pi
\over 2s}$, 2s being an integer, we show that the resulting
model is exactly
equivalent to an interacting theory of spin-$s$ fields. Thus we conjecture that
 there should be a fixed point in the spin-$s$ theory near which it is
 exactly equal to the $\sigma$ model.
\end{abstract}
\narrowtext
Interest in the 2+1 dimensional $ O(3)$ nonlinear sigma model (NLSM)
  with the Hopf term  was roused when Wilczek and Zee \cite{WZ} pointed out
that the Hopf term
causes the solitons to acquire fractional spin and statistics. The
connection of the NLSM without the  Hopf term to the long wavelength
fluctuations of antiferromagnets had been established by Haldane \cite{H}.
Though
the Hopf term has not been derived from any microscopic spin model so far,
it is conceivable that the NLSM  with the Hopf term may be relevant to
some class of frustrated antiferromagnets \cite{WWZ}.

The nonlocal nature of the Hopf term makes the treatment of the NLSM
difficult. Dzyaloshinski, Polyakov and Wiegmann \cite{DPW} suggested that the
model
was better studied in the $CP_1$ formalism, where in the long wavelength
limit, the effect of the Hopf term is to add  a local
Chern-Simons term to the model. Polyakov\cite{P} then showed that relativistic
scalar
particles interacting with abelian Chern-Simons gauge fields ( with a
particular coefficient) are Dirac fermions in the long wavelength limit.
This can easily be generalised to higher integer and half odd integer
spins as well \cite{SH}. Some of us \cite{US} then showed that the technique
works independent
of the long wavelength limit and also in the presence of self
interactions\cite{PSS}.

In this paper we use the above techniques to argue that the NLSM  with the
Hopf term with the coefficient $\theta={\pi\over 2s}\,(s={1\over 2},1,{3\over
2}\cdots )$ is equivalent to an interacting spin $s$ theory. In particular
when $s={1\over 2}$, the theory is equivalent to Dirac
fermions with four fermi interactions. The equivalence of the
$\theta=\pi$ NLSM to a four fermi theory has been recently conjectured on
general grounds \cite{K}. The topological current of the NLSM
is equivalent to the particle current of the spin $s$ theory.

We begin with the euclidean action of the NLSM ,
\begin{equation}
S_{NLSM}=g^2\int d^3x~\partial_\mu n^a \partial_\mu n^a~+~i\theta H[{\bf
n}]\label{NLSM}
\end{equation}
Here ${\bf n.n }=1$ and $H[{\bf n}]$ is the Hopf
invariant which is defined in the standard way. We use the euclidean
metric with the signature $(+,+,+)$. The $CP_1$ formalism \cite{YA} uses the
fact
that $S^2=SU(2)/U(1)$, and expresses the ${\bf n} $ fields in terms of $SU(2)$
matrices defined by $U^\dagger\tau^3U=n^a\tau^a$, where $\tau^a$ are the
Pauli matrices.  There is then a local $U(1)$
gauge invariance $U\rightarrow e^{i\Omega \tau^3}U$ which ensures
that the physical configuration space is unchanged. The action can then be
written in terms of the currents $L^a_\mu \equiv {1\over 2}tr\,(\tau^a
i\partial_\mu
UU^\dagger )$ as follows
$$S_{CP_1}=\int_x\left(   4g^2(L_\mu^aL_\mu^a-L^3_\mu L^3_\mu )-{{i\theta
}\over{4\pi^2}}
\in_{\mu\nu\lambda}L^3_\mu\partial_\nu L^3_\lambda\right) $$

Note that the action has been made local. Next we introduce two
auxilliary vector fields $a_\mu$ and $b_\mu$ as follows,
$$exp\left( 4g^2\int_x L^3_\mu L^3_\mu\right) =\int_{b_\mu} exp\left(
4g^2\int_x (2b_\mu L^3_\mu- b_\mu
b_\mu )\right)  $$
$$exp\left( {{i\theta}\over {4\pi^2}}\int_x  \in_{\mu\nu\lambda}L^3_\mu
\partial_\nu L^3_\lambda\right) =\int_{a_\mu}
exp\left( {{i\theta }\over{4\pi^2}}\int_x (2 \in_{\mu
\nu\lambda}a_\mu\partial_\nu L^3_\lambda - \in_{\mu\nu\lambda}a_\mu
\partial_\nu a_\lambda )\right) $$
Now note that
the NLSM  is periodic in $\theta$, namely $\theta$ and $\theta+2n\pi$
define the same theory. This is because the Hopf invariant is integer
valued. In terms of the $a_\mu$ fields, the periodicity is not
manifest. However if we make the transformation
$a_\mu\rightarrow ca_\mu+(1-c)L^3_\mu$, where
$c=(1+{{2n\pi}\over{\theta }})^{{1\over 2}}$,then that has the effect of
replacing $\theta $ by $\theta +2n\pi$ in the action for the $a_\mu$
field. Thus the periodicity is not spoilt.

We now parametrize the $SU(2)$ matrices by two complex fields
$z_\sigma,(\sigma=1,2)$ which satisfy the constraint ${\bf
z^\dagger z} =4g^2$ as
 \begin{eqnarray*}U\;&=&\;\frac{1}{2g}\left(
 \begin{array}{cc}
z_1&z_2\\-z^*_2&z^*_1
\end{array}\;\;\right)\;
\end{eqnarray*}
.The action can then be written as,
\begin{equation}
S_z = \int_x {\bf z}^\dagger{\cal D}{\bf z} +S_\eta +S_{GF}\label{SZ}
\end{equation}
where ${\cal D}= D_\mu
D_\mu+2i\sqrt\lambda\eta ,~~ D_\mu=i\partial_\mu-A_\mu ,~~ A_\mu=b_\mu+i\alpha
\in_{\mu\nu\lambda}\partial_\nu a_\lambda$ and~~ $\alpha =
{\theta\over{16\pi^2g^2}}$.
 The action for the
gauge fields is given by
\begin{equation}
S_{GF}=4g^2\int_x (-2i\alpha \in_{\mu\nu\lambda}b_\mu\partial_\nu
a_\lambda
+\alpha^2(\in_{\mu\nu\lambda}\partial_\nu a_\lambda )^2+i\alpha
\in_{\mu\nu\lambda}a_\mu\partial_\nu a_\lambda)
\end{equation}
The $\eta$ field is an auxilliary field which linearizes the term
$\lambda({\bf z^\dagger z} -4g^2)^2 $ in the action. This term
enforces the constraint ${\bf z^\dagger z}=4g^2 $, when
$\lambda \rightarrow \infty $.$S_\eta =\int_x (\eta^2-8i\sqrt\lambda
\eta  g^2 )$

We will from now on work with the model defined in Eq.\ref{SZ} which we will
refer to
as the $z$ theory. What we have shown above, is that the $z$ theory is
formally exactly equivalent to the NLSM defined in  Eq.\ref{NLSM}
in the limit of
$\lambda\rightarrow\infty$.

First we analyse the partition function of the model. The {\bf z}
fields can be integrated out and we have
\begin{equation}
{\cal Z}= \int_{a_\mu, b_\mu ,\eta}exp (-2~lndet{\cal D}-S_{GF}-S_\eta)
\end{equation}

Using the heat kernel representation of the logarithm of the determinant,
it can be written as a path integral as follows,
\begin{equation}
-2\; ln~det {\cal D} = 2\int_{1\over {\Lambda^2}} ^\infty {{d\beta }\over\beta}
\int_{x(\tau)}exp\left( -\int_0^\beta d\tau ({1\over {4}} (\partial_\tau x^\mu
)^2
+V(x))-i\oint A_\mu dx^\mu \right) \label{PI}
\end{equation}
where we have put $V(X)\equiv 2i\sqrt\lambda\eta$ , and  $\Lambda$ is the
ultraviolet cut off. The above two equations show that the dependence on
the gauge field $A_\mu$ can be expressed completely in terms of products
of Wilson loops.The averaging over gauge fields of an arbitrary product
of Wilson loops can be done as follows. If $C_i\; i=1\cdots n$ , denote n
distinct loops and $C\equiv \bigcup C_i$ , then we have
\widetext
\begin{eqnarray*}&&\int_{b_\mu} exp\left( -i\oint_C A_\mu dx^\mu
\right) exp\left( -S_{GF}[a,b]\right)=\delta(j_\mu^C(x)-{\theta\over
2\pi^2}\in_{\mu\nu\lambda}
\partial_\nu a_\lambda (x))\hfill\\ &&\hfill
exp\left(\int_x\left(\alpha(\in_{\mu\nu\lambda}\partial_\nu
a_\lambda)j^C_\mu(x)-4g^2\alpha^2(\in_{\mu\nu\lambda}\partial_\nu a_\lambda
)^2-4ig^2\alpha \in_{\mu\nu\lambda}a_\mu\partial_\nu a_\lambda\right)\right)
\end{eqnarray*}
\narrowtext
where
$$j_{\mu}^{C}(x)=
\sum_{i=1}^{n} \int_{0}^{\beta} \partial_{\tau} x_{\mu}^{C_{i}}(\tau)
\delta^{3}(x-x_{\mu}^{C_{i}}(\tau))$$
$ x_{\mu}^{C_{i}}(\tau)$ is the curve
$C_{i}$. It is now obvious that the $a_{\mu}$ integrals can be done easily.
However note  that at $\theta=0$, we get $j_{\mu}^{C}=0$. This then
implies that every $z$ particle is accompanied by an antiparticle. Thus
single $z$ particles cannot propagate and are hence confined. The presence
of the Hopf term thus leads to the possibility of the $z$ particles being
deconfined much in the way deconfinement occurs in Chern-Simons gauge
theories coupled to matter as stated in reference \cite{DPW}. The above
discussion also shows that the limit of $\theta \rightarrow 0$ i.e. $s
\rightarrow \infty$ may not be smooth and in particular may not be the
$\theta=0$ theory.

When the $a_{\mu}$ integrals are done, we obtain in the exponential,
$$
-~{{i\pi^2}\over\theta}~{1\over{4\pi}}\oint_C dx^\mu\oint_C
dy^\nu\in_{\mu\nu\lambda}{{(x-y)_\lambda}\over{|x-y|^3}}~+~
{1\over 16g^2}\int_xj_\mu^C(x)j_\mu^C(x)
$$
The first term is ${i\pi^2 \over \theta}(\sum_{i=1}^{n} W[C_{i}] +
\sum_{i \neq j} 2n_{ij}^{l})$, where  $W[C_{i}]$ is the writhe of the
curve \cite{FKV} $C_{i}$ and $n_{ij}^{l}$ is the linking number of the curves
$C_{i}$ and $C_{j}$. Since $\theta={\pi \over 2s}$ we can see that
the exponential of the linking number term is unity. We then introduce
an auxilliary vector field $v_{\mu}$ to linearize the second term and
obtain the general identity,
$$
\int_{a_\mu,b_\mu}e^{-S_{GF}}F[\prod_{i=1}^ne^{-i\oint_{C_i}A_\mu dx^\mu}]~
=~\int_{V_\mu}e^{-\int_x v_\mu v_\mu}F[\prod_{i=1}^ne^{-{1\over
2g}\oint_{C_i}v_\mu dx^\mu-2\pi isW[C_i]}]
$$
 Note that $W[C_{i}]={1 \over
 2\pi}\Omega[C_{i}] + 2k+1$, where  $\Omega[C_{i}]$ is the solid angle
 subtended by the curve on the $2$-sphere traced out by the unit tangent
 to $C_{i}$ and $(2k+1)$ is some odd integer. Thus $\exp (-2\pi isW[C_{i}])=
 (-1)^{2s}\exp(-is\Omega[C_{i}])$.

 Using this identity, it can be seen that after the gauge field
 integrations, the partition function can be written as,
$$
{\cal Z}~=~\int_{\eta,v_\mu}e^{-S_\eta-\int_xv_\mu v_\mu -2P}
$$
 where $P$ is the path integral in Eq.\ref{PI}, with the Wilson loop
 term replaced by $(-1)^{2s}\exp(-is\Omega[C])$. To do this path integral, we
 introduce velocity variables $u_{\mu}=\partial_{\tau}x_{\mu}(\tau)$ and
 lagrange multipliers $k_{\mu}$ to impose the constraint. The path
 integral is then written as,
$$
(-1)^{2s}~2~\int_{x(\tau)u(\tau)k(\tau)}
exp\left( -\int_0^\beta d\tau(ik_\mu \dot x^\mu+{1\over 4}(u_\mu)^2+V(x)-ik_\mu
u^\mu+is\Omega[{\bf u}])\right)
$$

 This expression can be shown, as detailed in refs. \cite{PSS} to be equal
 to $(-1)^{2s}Tr\exp(-L{\cal D}^{(s)})$. Here ${\cal D}^{(s)}=(i\partial_\mu
+i{1
 \over 2g}v_{\mu}){T^{\mu} \over s}+M_{s}+\kappa V(x)$. $T^{\mu}$ are the
 generators of $SU(2)$ in the spin-$s$ representation. $L={1 \over
 \kappa}\beta, M_{s}=\Lambda^{2}\kappa\ln(2s+1)$ and $ \kappa={\sqrt{\pi}
 \over 4\Lambda}$. The best way to see this equality is to start from
 $Tr~\exp(-LD^{(s)})$ and derive the path integral for it using a
(over)complete
 set of states $\mid {\bf x}\rangle\mid{\bf u}\rangle$ at every $\tau$ slice,
 $\mid {\bf u}\rangle$ being the spin-$s$ coherent states. One then obtains
 the path integral above after doing the radial u integrals. In this
 calculation, we have put $\epsilon$, the interval between two $\tau$
 slices, to be equal to ${1 \over \Lambda^{2}}$.

 Doing the $L$ integral, we obtain $P=(-1)^{2s}~2~\ln detD^{s}$. The
 determinant can be written as a path integral over two complex fields
 $ \bar\psi_{m\sigma}, \psi_{m\sigma} (\sigma=1,2\; m=-s\cdots s)$
 transforming under the spin-$s$ representation. These are Grassmann
 fields for $2s=odd$ and bosonic fields for $2s=even$. The $v_{\mu}$ and
 $\eta$ intgrals can also be done and we finally obtain for the
 partition function,
\begin{equation}
{\cal
Z}~=~\int_{\bar\psi_\sigma,\psi_\sigma}exp\left( -\int_x(\bar\psi({T^\mu\over
s}
i\partial_\mu+M)\psi+\gamma_1(\bar\psi{T^\mu\over s}\psi)^2
+\gamma_2(\bar\psi\psi)^2)\right)\label{HS}
\end{equation}
where, $M=M_s-8g^2\kappa\lambda$, $\gamma_1={1\over 16g^2}$,
$\gamma_2=\lambda\kappa$
 Thus the partition function of the $z$ theory is equal to that of the
 above spin-$s$ theory. Note by spin-$s$, we refer to the representation of
 the full Lorentz group and not just rotations. The fields in this
 theory satisfy the 2+1 dimensional Bhabha equations. They, in general
 describe a multiplet of particles with different masses.

Next, by coupling the topological current to an external gauge field, we
can derive, using the same procedure, the fact that, $\langle
j_{\mu}^{top}(x)\rangle_{z}=i{\pi \over \theta}\langle \bar\psi(x)
{T^{\mu} \over s}\psi(x)\rangle_{s}$. The subscripts refer to the theory in
which the averaging is done. Thus it follows that the average number of
solitons $N_{top}=2sN_{\psi}$ where $N_\psi$ is the number of $\psi$ particles
(the
relative factor of $i$ goes away when we continue to Minkowski space).
Thus a single $\psi$ particle corresponds to a soliton number $2s$ object
on the average. The relations between the current-current correlation
functions can also be derived and we obtain,
$$
\langle j_\mu^{top}(x)j_\nu^{top}(y)\rangle~=~-\left({\pi\over
\theta}\right)^2\langle
J_\mu^s(x)J_\nu^s(y)\rangle~+~{i\over 2\theta}\in_{\mu\nu\lambda}\partial
_\lambda\delta^3(x-y)
$$
Note that the second term on the RHS is not parity invariant. However
neither is the first term. For the case $s={1 \over 2}$, the $\psi$
theory is that of Dirac fermions. The parity noninvariant part of the
correlator comes from the induced Chern-Simons term and the coefficient
is such that it exactly cancels the second term. This does not happen in
the higher spin theories. Thus the soliton current-current correlator is
parity invariant only at $\theta=\pi$ as we expect from the NLSM.

We now address the question of what corresponds to the $\psi$ fields in
the $z$ theory. We find, similar to reference \cite{US} that if we define,
$$
\chi_{\sigma{\bf u}
}(x)~=~\zeta(C)exp\left({i\int_C^xL_\mu^3dx^\mu}\right)\Delta(u_\mu-\in_{\mu\nu\lambda}
\partial_\nu L^3_\lambda)z_\sigma(x)
$$
$$
\bar\chi_{\sigma{\bf u}}(x)~=~\bar\zeta(\bar C)exp\left({-i\int_{\bar
C}^xL^3_\mu dx^\mu}\right)
\Delta(u_\mu-\in_{\mu\nu\lambda}
\partial_\nu L^3_\lambda)z^*_\sigma(x)
$$
Then we obtain $\langle\chi_{\sigma{\bf u}}(x)\bar\chi_{\sigma'{\bf u'}
}(x')\rangle_{z}=\langle\psi_{\sigma {\bf u}}(x)\bar\psi_{\sigma'{\bf u'}
}(x')\rangle_{s}$. Here $\psi_{\bf u}\equiv \langle{\bf u}
\mid\psi\rangle$, where $\mid{\bf u}\rangle$ is the spin-$s$ coherent
state representing a spin, polarized in the direction ${\bf u}$,
$\Delta(u_{\mu}-
\in_{\mu\nu\lambda}
\partial_\nu L^3_\lambda)$ is defined as
$\int_{0}^{\infty}duu^{2}\delta^{3}(u_{\mu}-\in_{\mu\nu\lambda}
\partial_\nu L^3_\lambda)$.~$ \zeta[C]={1\over\sqrt{\kappa}}
exp\{-is\Omega[C,u^*]\}$~and~$\bar\zeta[\bar C]={1\over\sqrt{\kappa}}
exp\{is\Omega[\bar C,u^*]\}$. $C$~and~$\bar C$ are some curves from infinity
to $x$. $\omega[C,u^*]$ for open curves on the two sphere is defined as
the solid angle of the closed curve obtained by connecting the two end
points to a standard point,~$u^*$~by geodesics. Thus
the Lorentz components of the $\psi$ field at the point $x$ are determined
by the direction of the topological current at $x$. The details of the
above calculations are being presented elsewhere\cite{UP}.

The transformation of the $\psi$ fields under (euclidean) Lorentz
rotations are manifest. The topological current occuring in the
definition of the $\chi$ fields transforms like a vector. This leads to the
spin
coherent states and hence the $\psi$ fields transforming in the spin-$s$
representation. Discrete transformations however are more subtle. In the path
integral, the
term $k_\mu u_\mu$ is what becomes the term, ${T^\mu \over
s}i\partial_\mu$ in the spin-$s$ operator. Since both $u_\mu$ and
$k_\mu$ change sign under parity, it would seem that the operator and
hence the $\psi$ fields are scalars under parity. However,
the $z$ theory is not parity
invariant because the Chern-Simons term changes sign. This leads to an
inversion in the definition of the Lorentz spin axes in the coherent
states. We are able to then show that consequently, ${T^{\mu} \over
s}i\partial_{\mu} \rightarrow -{T^{\mu} \over s}i\partial_{\mu}$ in the
spin-$s$ operator.
Thus the $\psi$ fields transform in the standard way
under parity. Similar calculations  show that a charge
conjugation of the $\chi$ fields leads to the standard charge conjugation of
the
$\psi$ fields.
The phase factors in $\zeta[C]$ have to be carefully dealt with in this
calculation. As a consequence of this, though under parity {\bf
z}$^\dagger${\bf z} is a
scalar, $\bar\psi\psi$ is a pseudo scalar. The details of these
calculations  will be presented elsewhere
\cite{UP}.

In conclusion, what we have shown is that the $z$ theory is exactly
equivalent to the spin-$s$ theory in Eq.\ref{HS}. The $z$ theory is
formally equivalent to the NLSM in Eq.\ref{NLSM} when
$\lambda\rightarrow\infty$. Further, the NLSM has been shown to be a
renormalizable theory \cite{Aa}. So we expect, as in the $\theta=0$ theory
\cite{RG},
that there should be a $\lambda=\infty$ fixed point of the spin-$s$ theory
near which it will reproduce the correlation functions of the NLSM. Note
that at $s={1 \over 2}$, the NLSM is parity invariant whereas the
fermionic theory is not. However this is not surprising since the $z$
theory is also not parity invariant for finite $\lambda$. Thus we expect
that parity should be recovered near the fixed point and hence the
continuum fermions should be massless.

An application of this formalism could be to anyon
superconductivity in the NLSM, which
 for general $\theta$ contains anyonic solitons. If we
couple the soliton current to an external electromagnetic field, then it
will couple to the particle Noether current in the spin-$s$ theory. Hence
we could discuss the superconductivity of anyons
in conventional terms, i.e. in terms of $U(1)$ symmetry breaking and
order parameters for the same.

\acknowledgements
NS and MS would like to thank The Institute of Mathematical Sciences,
Madras, India for hospitality. RS and TRG thank R. Shankar(Yale
University) for useful discussions.


\begin{references}
\bibitem{WZ} F.Wilczek and A.Zee, Phys.\ Rev.\ Lett.\ {\bf 50}, 2250
(1983).
\bibitem {H} F.D.M.Haldane, Phys.\ Rev.\ Lett.\ {\bf 50}, 1153 (1983).
\bibitem {WWZ} X.G.Wen, F.Wilczek and A.Zee, Phys.\ Rev.\ B{\bf 39},
11413 (1989).
\bibitem {DPW} I.Dzyaloshinskii, A.M.Polyakov and P.B.Wiegmann, Phys.\
Lett.\ {\bf 127}A, 112 (1988).
\bibitem {P} A.M.Polyakov, Mod.\ Phys.\ Lett.\ A{\bf 3}, 325 (1988).
\bibitem {SH} A.Yu.Alekseev and S.L.Shatashvili, Mod.\ Phys.\ Lett.\ A{\bf
3}, 1551 (1988).
\bibitem {US} N.Shaji, R.Shankar and M.Sivakumar, Mod.\ Phys.\ Lett.\
A{\bf 5}, 593 (1990); S.Ito, C.Itoi and H.Mukaida, Nucl.\ Phys.\ B{\bf
346}, 293 (1990).
\bibitem {PSS} S.K.Paul, R.Shankar and M.Sivakumar, Mod.\ Phys.\ Lett.\
A{\bf 6}, 553 (1991); R.Shankar and M.Sivakumar, Mod.\ Phys.\ Lett.\
A{\bf 6}, 2379 (1991).
\bibitem {K} A.Kovner, Int.\ J.\ Mod.\ Phys.\ A{\bf 5}, 3999 (1990).
\bibitem {YA} Y.S.Wu and A.Zee, Phys.\ Lett.{\bf 147}B,325(1984)
\bibitem {FKV}M.D.Frank-Kamenetskij and A.V.Vologodskij, Sov.\ Phys.\
Usp.\ {\bf 24}, 679 (1981).
\newpage
\bibitem {UP} T.R.Govindarajan, N.Shaji, R.Shankar and M.Sivakumar,
under preparation.
\bibitem {Aa} I.Ya.Arefyeva, Ann. of Phys.(N.Y){\bf 117},(1979)393.
\bibitem {RG} J.Zinn-Justin,``Quantum Field Theory and Critical
Phenomena", Clarendon Press, Oxford (1989); P.K.Mitter and T.R.Ramdas,
Comm.\ Math.\ Phys. {\bf 122},575 (1989).
\end{references}
\end{document}